\documentclass[runningheads,a4paper]{llncs}

\usepackage{cmap}
\usepackage[T1]{fontenc}
\usepackage{graphicx}
\usepackage[ngerman,english]{babel}
\addto\extrasenglish{\languageshorthands{ngerman}\useshorthands{"}}
\usepackage[%
rm={oldstyle=false,proportional=true},%
sf={oldstyle=false,proportional=true},%
tt={oldstyle=false,proportional=true,variable=true},%
qt=false%
]{cfr-lm}
\usepackage[math]{blindtext}
\usepackage{cite}
\usepackage{paralist}
\usepackage{csquotes}
\usepackage{microtype}
\usepackage{url}
\makeatletter
\g@addto@macro{\UrlBreaks}{\UrlOrds}
\makeatother
\usepackage{xcolor}
\usepackage{pdfcomment}
\hypersetup{hidelinks,
   colorlinks=true,
   allcolors=black,
   pdfstartview=Fit,
   breaklinks=true}
\usepackage[all]{hypcap}

\usepackage[capitalise,nameinlink]{cleveref}
\crefname{section}{Sect.}{Sect.}
\Crefname{section}{Section}{Sections}
\usepackage{xspace}

\DeclareFontFamily{U}{MnSymbolC}{}
\DeclareSymbolFont{MnSyC}{U}{MnSymbolC}{m}{n}
\DeclareFontShape{U}{MnSymbolC}{m}{n}{
    <-6>  MnSymbolC5
   <6-7>  MnSymbolC6
   <7-8>  MnSymbolC7
   <8-9>  MnSymbolC8
   <9-10> MnSymbolC9
  <10-12> MnSymbolC10
  <12->   MnSymbolC12%
}{}
\DeclareMathSymbol{\powerset}{\mathord}{MnSyC}{180}
\usepackage{tabularx}
\newcolumntype{L}[1]{>{\raggedright\arraybackslash}p{#1}} 
\newcolumntype{C}[1]{>{\centering\arraybackslash}p{#1}} 
\newcolumntype{R}[1]{>{\raggedleft\arraybackslash}p{#1}} 
\usepackage{multirow}
\usepackage{pifont}
\usepackage{colortbl}
\usepackage{hhline}
\usepackage{verbatim}

\begin{document}

\title{Demystifying Deception Technology:\\A Survey}

\titlerunning{Demystifying Deception Technology}

\author{
Daniel Fraunholz\inst{1} \and Simon Duque Anton\inst{1} \and Christoph Lipps\inst{1} \and Daniel Reti\inst{1} \and Daniel Krohmer\inst{1} \and Frederic Pohl\inst{1} \and Matthias Tammen\inst{1} \and Hans Dieter Schotten\inst{1,2}}
\authorrunning{Fraunholz et al.}

\institute{
Intelligent Networks Research Group \\German Research Center for Artificial Intelligence\\Kaiserslautern, Germany\\
\email{\{firstname\}.\{lastname\}@dfki.de}\and
Institute for Wireless Communication and Navigation\\ University of Kaiserslautern\\Kaiserslautern, Germany\\
\email{schotten@eit.uni-kl.de}
}

\maketitle

\begin{abstract}
Deception boosts security for systems and components by denial,
deceit,
misinformation,
camouflage and obfuscation.
In this work an extensive overview of the deception technology environment is presented.
Taxonomies,
theoretical backgrounds,
psychological aspects as well as concepts,
implementations,
legal aspects and ethics are discussed and compared.
\end{abstract}


\section{Introduction}\label{sec:intro}
\textit{Sun Tzu} once wrote ``all warfare is based on deception'' \cite{Sun.2001}.
This was long before the first digital devices.
Since then,
2500 years ago,
deception was an essential aspect of many fields,
e.g. the military.
Over the years deception was an essential aspect of military operations.
In \textit{information security} (IS),
social engineering,
as extensively described by \textit{Mitnick} \cite{Mitnick.2002} was the first use of deception.
In 1986 and 1991,
\textit{Stoll} \cite{Stoll.1989},
respectively \textit{Cheswick} \cite{Cheswick.1991},
transferred the concept of deception to defensive applications.
These applications were called \textit{honeypots} (HP).
Later on the concept was generalized to \textit{Deception technology} (DT),
which is a superset of \textit{HP}s and all other technologies relying on simulation and dissimulation.
The \textit{Deception Toolkit} published by \textit{Cohen} was the first publicity available deception software \cite{Cohen.1995}. 
Over the last three decades,
the concepts of deception experienced a rising popularity in information security.
Perimeter-based security measures,
such as firewalls and
authentication mechanisms,
do not provide a proper level of security in the context of insider threats and social engineering.
Defense-in-depth strategies,
such as signature-based intrusion detection and prevention,
often suffer from a large number of false-positive detections,
resulting in alarm fatigue of security information and event management systems.
\textit{DT}s come along with several major advantages such as low false-positive alerts and 0-day detection capabilities,
making it a promising solution to tackle advanced security threats such as intrusion detection \cite{DuqueAnton.2017b} and attribution \cite{Fraunholz.2017e,Fraunholz.2017c}.
This work is structured as follows:
In section \ref{sec:background},
an overview of recent definitions and taxonomies related to \textit{DT} are discussed and a taxonomy for this work is specified.
\textit{DT} is also put into context of the current IT-security environment.
Furthermore,
cognitive vulnerabilities are introduced and formal deception models are reviewed.
Section \ref{sec:software} gives a comparative overview of important and recent advances in deceptive software,
\textit{honeytokens} (HT) and \textit{HP}s as well as field studies based on such technologies.
Legal considerations, ethics and baseline security is discussed in section \ref{sec:legal}.
In section \ref{sec:conclusion},
this work is concluded.

\section{Background and Theory} \label{sec:background}

In this section,
research on the theoretical background of deception is reviewed.
First,
definitions and taxonomies are presented and determined for this work.
\textit{DT} is then integrated in the information security environment and research on psychological aspects of deception is discussed.
The section is finished by a review on formal approaches to model deception as a game.

\subsection{Definitions and Taxonomies}

\textit{Whaley} \cite{Whaley.2008} defines deception as a misperception that is intentionally induced by other entities.
In his typology of perception,
deception has three requirements:
1) not a pluperception,
2) not self-induced and 3) not induced unintentional.
Around ten years after this publication \textit{Bell and Whaley} \cite{Bell.1991} published a taxonomy of deception.
This taxonomy classifies deception into two major categories: dissimulation and simulation.
Dissimulation consists of three classes: masking,
repacking and dazzling,
whereas mimicking,
inventing and decoying are within the simulation category.
This taxonomy is the most frequently employed taxonomy for deception in the information security domain.
In this work,
the taxonomy from \textit{Bell and Whaley} is used.
\textit{Dunnigan and Nofi} \cite{Dunnigan.op.2001} categorized deception based on the six principles of deception introduced by \textit{Fowler and Nesbit} \cite{Fowler.1995}: 
Deception should reinforce enemy expectations,
have realistic timing and duration,
be integrated with operation,
be coordinated with concealment of true intentions,
be tailored to needs of the setting and be imaginative and creative.
Their taxonomy differentiates nine classes of deception:
Concealment,
camouflage,
false and planted information,
lies,
displays,
ruses,
demonstrations,
feints and insight.
\textit{Rowe and Rothstein} \cite{Rowe.2004} proposed a taxonomy based on the works of \textit{Fillmore} \cite{Fillmore.1992},
\textit{Copeck} \cite{Copeck.1992} and \textit{Austin} \cite{Austin.ca.2009} on linguistic cases.
Their taxonomy consists of 32 cases in 7 groups.
\textit{Stech et al.} \cite{Stech.2011} found that deception is defined in literature in a number of ways.
They published a scientometric analysis of the concept of deception in the cyber-space domain.
\textit{Monroe} \cite{Monroe.2012} compared several competing definitions of deception.
He introduces a taxonomy that closely matches \textit{U.S. Army} doctrinal concepts.
The taxonomy is structured in three layers,
where the two main categories (active deception and cover) comprise 13 atomic peculiarities of deception.
In 2012 the \textit{U.S. Army} \cite{U.S.Army.2014} published a document that grouped \textit{MILDEC} as part of information operation and related to \textit{electronical warfare} (EW),
\textit{psychological operation} (PSYOPS),
\textit{computer network operation} (CNO) and \textit{operation security} (OPSEC).
\textit{Shim and Arkin} \cite{Shim.2013} compared several taxonomies for deception in respect to the domain.
In the information security domain,
the previously introduced taxonomy from \textit{Rowe} based on linguistic cases was chosen.
Other domains are: Philosophy,
psychology,
economics,
military (taxonomy from \textit{Bell and Whaley}) and biology.
\textit{Almeshekah and Spafford} \cite{Almeshekah.2014} published a work on planning and integrating of deception.
They proposed a taxonomy with focus on the deception targets.
Later on,
\textit{Almeshekah} \cite{Almeshekah.2015} extended the previous work by grouping deception and other security mechanisms into four categories.
He then identified intersections of the categories and mapped the mechanisms to the famous kill chain proposed by \textit{Hutchins et al.} \cite{Hutchins.2011}.
\textit{Pawlick et al.} \cite{Pawlick.2017} created a taxonomy by grouping existing game-theoretical approaches to deception as discussed later in this section. Their taxonomy divides deception into: Perturbation,
\textit{moving target defense} (MTD),
obfuscation,
mixing,
honey-x and attacker engagement.

More specific taxonomies for example for \textit{HP}s are given by \textit{Seifert et al.} \cite{Seifert.2006} or \textit{Pouget et al.} \cite{Pouget.2003}.
The term \textit{Deception Technology} recently became famous and is currently used for deception-based security that is superior to the 30-year old \textit{HP} technology.
It is frequently associated with holistic frameworks that comprise:
Adaptive \textit{HT} and decoy generation,
automated deployment and monitoring as well as integrated visualization and reporting capabilities.

\subsection{Deception in the IT-Security Environment}
\begin{figure}
\centering
\includegraphics[width=1\textwidth]{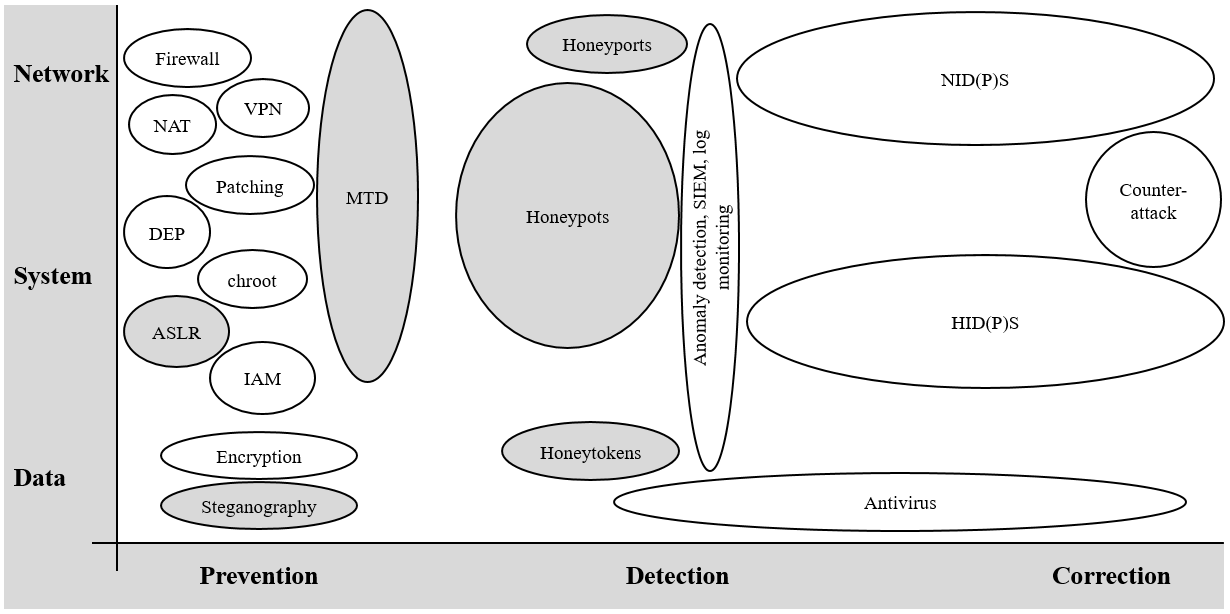}
\caption{Overview of relevant security techniques and primitives on different layers, grey: deception-based technologies}
\label{fig:itsec-measures}
\end{figure}

The term \textit{IT security} summarizes a broad range of technical and organizational measures aiming for the protection of a predefined set of assets.
These assets can be classified according to the so-called \textit{CIAA} security goals.
\textit{CIAA} stands for \textit{Confidentiality, Integrity, Authentication and Availability} and describes the protection goals of common \textit{IT}-systems.
\textit{DT} cannot be mapped onto those security targets,
but is commonly used for more abstract objectives,
e.g. intrusion detection or analyzing attacker behavior.
Another fundamental difference between classic \textit{IT security} and \textit{DT} is the use of \textit{Security by Obscurity}.
This term describes the intentional use of methods more complex than necessary or not published in order to prevent an attacker from gaining knowledge about the system.
While it is not advised,
and has repeatedly backfired in classic \textit{IT security},
it is a valid technique to slow down attackers or distract them from possible targets.
\textit{DT} does not necessarily depend on obscurity,
but is significantly more effective if the presence is unknown.
The relation of classic \textit{IT security} and \textit{DT} is shown in figure~\ref{fig:itsec-measures}.

In this figure,
three different layers of resources to protect,
as well as three different protection mechanisms are shown.
The resources can be grouped into network-,
system and data-resources.
Protection can be done by preventing an attack,
by detecting it and by responding to it in a way that hinders the result of it,
so-called correction.
There are a lot of classic \textit{IT security} building blocks that deal mostly with prevention,
such as \textit{Virtual Private Networks (VPN)},
\textit{Firewalls},
\textit{Data Execution Prevention (DEP)},
\textit{Identity and Access Management (IAM)} and encryption.
A few of the well-established mechanisms,
such as \textit{(Kernel) Address Space Layout Randomization ((K)ASLR)} or steganography,
can be considered as \textit{DTs} according to the taxonomies as presented in the previous subsection,
as they harden systems and protect resources by making it harder for an attacker to find an attack surface.
A few more novel \textit{DTs} address prevention,
such as \textit{MTD},
\textit{Random Host Mutation (RHM)},
\textit{Dynamic Instruction Set (DIS)} and \textit{Unified Architecture (UA)}.
More well-known \textit{DTs},
such as \textit{HPs} and \textit{HTs},
address detection.
This area is furthermore only addressed by a few classic \textit{IT security}-mechanisms,
such as \textit{Network and Host Intrusion Detection (and Prevention) Systems (N/HID(P)S)},
\textit{Security Information and Event Management Systems (SIEMs)} and antivirus-software.
More novel technologies,
such as anomaly detection algorithms,
address this protection goal as well,
with similar methodologies as \textit{DTs}.
According to \textit{Corey},
\textit{HP}s,
in their very essence,
act as a anomaly-based intrusion detection system~\cite{Corey.2003}.
Even though there is an overlap,
this statement is only true for the detection mechanism,
as all incoming traffic can be considered an anomaly and can automatically be classified as an attack.
In general the characteristic of an \textit{HP},
as well as \textit{DT}s in general,
is to send misleading and wrong data,
whereas anomaly detection is only receiving and analyzing information.

\subsection{Cognitive Vulnerabilities}
Deception as defined by \textit{Whaley} \cite{Whaley.2008} as well as other authors,
strongly relies on human psychology.
In 1994 \textit{Libicki} \cite{Libicki.2004} coined the term \textit{semantic attacks}.
He distinguished it from \textit{physical} and \textit{syntactic attacks} and discussed it in the context of \textit{IS}.
Later on,
18 relevant propositions for deception in the context of general communication were identified and described by \textit{Buller and Burgoon} \cite{Buller.1996}.
They also investigated relevant factors for deception.
6 years later,
\textit{Cybenko et al.} \cite{Cybenko.2002} introduced the term cognitive hacking and proposed several counter measures such as source authentication,
trajectory modeling and \textit{Ulam} games.
The term cognitive hacking is mostly referred to in the context of deception in the news or social media.
\textit{Mitnick} \cite{Mitnick.2002} published an extensive work on social engineering,
which he defined as the use of influence and persuasion to deceive people.
Until today,
his work is often referred to as the definitive book for deception,
even though it only focuses on social engineering.
The automated training of techniques to detect deception was investigated by \textit{Cao et al.} \cite{Cao.2003}.
\textit{Cranor and Simson} \cite{Cranor.2008} stated that the security of any sociotechnical system is based on three elements:
product,
process and panorama.
In their work,
they analyzed the correlation of usability and security.
The impact of design on deception was further analyzed by \textit{Yuill} \cite{Yuill.2007}.
\textit{Ramsbrock et al.} \cite{Daniel.2007} conducted an empirical study to invstigate the behaviour of intruders after they were granted access to a honeypot.
This idea was later on employed by \textit{Sobesto} \cite{Sobesto.2015} as well as by \textit{Howell} \cite{ChristianJordanMichaelHowell.2011} to investigate the impact of design characteristics such as the welcome banner or system resources on the behavior of attackers in empirical studies.
A more systematic approach to identify a set of relevant cognitive biases was proposed by \textit{Fraunholz et al.}  \cite{Fraunholz.2017}.
They analyzed cognitive biases such as those introduced by \textit{Kahneman} \cite{Kahneman.2008},
aiming to match biases and design characteristics in order to finally derive basic design principles.

\subsection{Formal Deception Models}

Several researchers targeted the formal description of deception in games and models.
As introduced in a previous subsection,
\textit{Pawlick et al.} \cite{Pawlick.2017} recently published a game-theoretical taxonomy for deception.
In their work,
they also reviewed and categorized 24 publications of deception games.
Two works that were published after their review are discussed in this subsection.
\textit{Wang et al.} \cite{Wang.2017} published a deception game for a specific use case.
They focus on smart grid applications and the security benefits of deception against distributed denial of service attacks.
In 2018,
\textit{Fraunholz and Schotten} \cite{Fraunholz.2018} published a game which allows to model probes within the game.
An attacker as well as a defender are able to choose the effort they want to invest in the obfuscation of the deception systems or the examination of systems of unknown nature.
After the strategy is chosen,
the attacker decides whether to attack, probe or ignore the system.
They provided a heuristic solution for their game.

\section{Deception Technology and Implementations} \label{sec:software}
In this section three noteworthy topics are discussed.
First,
an overview of research on deceptive software and \textit{HT}s is given.
This overview excludes \textit{HP}s,
as they are reviewed in the second subsection.
Finally,
an overview of field studies and technological surveys on \textit{DT}s is given.

\subsection{Deceptive Software and Honeytokens}
The term \textit{Honeytoken} has been coined in 2007 by \textit{de Barros} \cite{Barros.2007}.
The idea can be traced back to \textit{Spitzner} \cite{Spitzner.2003},
who first defined a \textit{HT} as the equivalent to a \textit{HP},
with the constraint of representing an entity different from a computer.
Since the inception of deception in information security,
a vast number of concepts have been published.
They cover different types of entities and focus on the generation of deceptive twins and their deployment.
Many of these entities were given specific names referencing to the term honeypot e.g. honeywords for fake passwords in a database.
The range of domains for \textit{HT}s can be distinguished by the categories Server,
Database,
Authentication and File.
As a further abstraction,
\textit{HT}s can be issued the labels host-based (H) or network-based (N).
Both distinctions have been applied for table \ref{tab_deceptive_soft},
which gives an overview of recent deception and \textit{HT} techniques.
An effective deception to protect files is giving files fake metadata,
such as false author,
creation and modification date and file size.
The latter has been approached by \textit{Spafford} \cite{Spafford.2011},
using the \textit{Unix Sparse File} structure to deceit copy applications.
\textit{Rowe} \cite{Rowe.2007} presented the concept of making targets appear like \textit{HP}s,
taking advantage of the attackers fear of having their techniques exposed and resources wasted.
The deceptive elements he suggests for fake \textit{HP}s are the usage of \textit{HP} tools,
mysterious processes,
nonstandard system calls in key security subroutines,
dropping and modification of packets and manipulation of metadata to make it seem abandoned.
He also anticipates that this will lead to a further level of deception by making \textit{HP}s appear like fake \textit{HP}s.
\textit{Laur{\'en et al.}} \cite{Lauren.2016} suggest system call manipulation by changing the system call numbers, while monitoring the original ones, which the attacker will possibly try to use.
\textit{Bercovitch} \cite{Bercovitch.2011} developed a technique of automatically generating database entries through a machine learning algorithm that learns from existing entries.
This \textit{HT} generator is protecting manually marked sensitive data by changing it prior to the learning routine.
\textit{Juels and Rivest} \cite{Juels.2013} coined the term \textit{honeyword} for the idea of assigning multiple false passwords together with the real password to an account to decrease the value of data breaches.
For the distinction between \textit{honeywords} and the real password,
a dedicated authentication system was proposed and named \textit{honeychecker}.
An idea to forward suspicious authentication attempts into a honey account was proposed by \textit{Almeshekah et al.} \cite{Almeshekah.2015}.
\textit{Lazarov} \cite{Lazarov.2016} analyzed malicious and ordinary behavior by publishing and tracking \textit{Google} spreadsheets that contained fake bank accounts and \textit{URL}s on paste sites.
The idea of patching a systems security vulnerability but make it appear unpatched was proposed by \textit{Araujo et al.} \cite{Araujo.2014} and termed \textit{honeypatch}.
When trying to exploit the patched vulnerability,
the attacker's connection is redirected to a \textit{HP}.
They suggest a wide scale adoption would reduce vulnerability probing and other attacking activities.
A similar approach named \textit{ghost patches} was made by \textit{Avery and Spafford} \cite{Avery.2017}.
They propose to prevent security patches from exposing vulnerabilities by inserting fake patches into actual security patches.
For the protection of web servers,
\textit{Fraunholz et al.} \cite{Fraunholz.2018b} proposed deceptive response messages to mitigate reconnaissance activities.
Their deceptive web server provides fake banners with a false server and operating system versions,
fake entries in the \textit{robot.txt} file,
dynamically injected honeylinks and false error responses on file access to mitigate brute-forcing attempts of the file system.
A deception technique called \textit{MTD} does not consist of fake entities,
but rather approaches the randomization and mutation of the network topology and host configuration.
\textit{Okhravi et al.} \cite{Okhravi.2014} distinguished \textit{MTD} into the categories dynamic networks,
platforms,
runtime environments,
software and data.
Commonly used \textit{MTD} techniques are \textit{ASLR},
Instruction Set Randomization and Code Sequence Randomization.
\textit{Al-Shaer et al.} \cite{AlShaer.2012} proposed a \textit{MTD} technique they call \textit{RHM},
allowing host randomization using \textit{DNS} without requiring changes in the network infrastructure.
\textit{Park et al.} \cite{Park.2018} combined the principles of \textit{MTD} to dynamically generate network topologies and the injection of decoys.

\begin{table}[]
\centering
\scriptsize
\caption{Overview of deceptive software techniques}
\label{tab_deceptive_soft}
\begin{tabular}{|l|l|l|l|l|l|}
\hline
\rowcolor{lightgray!50}\textbf{Technique} & \textbf{Deceptive Entity} & \textbf{Domain} & \textbf{H} & \textbf{N} & \textbf{Ref}\\
\hline
Fake Honeypot & Honeypot & Server & \ding{55}&\ding{51} & \cite{Rowe.2007}\\
\hline
Honeyentries & Table, data set & Database & \ding{51} & \ding{55} & \cite{Bercovitch.2011,Fraunholz.2018c}\cite{Hoglund.2011}\\
\hline
MTD & Topo., net. interf., memory, arch. & Versatile & \ding{51} & \ding{51} & \cite{Okhravi.2014,AlShaer.2012,Park.2018}\\
\hline
Honeyword & Password & Authentication & \ding{51} & \ding{55} & \cite{Juels.2013}\\
\hline
Honeyaccount & User account & Authentication & \ding{51} & \ding{55} & \cite{Almeshekah.2015,Fraunholz.2018c}\\
\hline
Honeyfile & (Cloud-)File & File system & \ding{51} & \ding{51} & \cite{Lazarov.2016,Fraunholz.2018c}\\
\hline
Honeypatch & Vulnerability & Server & \ding{51} & \ding{51} & \cite{Araujo.2014,Avery.2017}\\
\hline
- & Memory & Server  & \ding{51} & \ding{55} & \cite{Dowd.2007}\\
\hline
- & Metadata & File & \ding{51} & \ding{55} & \cite{Rowe.2007}\\
\hline
HoneyURL & URL & File & \ding{55}&\ding{51} & \cite{Lazarov.2016}\\
\hline
Honeymail & E-Mail adress & File & \ding{55} & \ding{51} & \cite{Rowe.2007b,Fraunholz.2018c}\\
\hline
Honeypeople & Social network profile & File & \ding{55} & \ding{55}& \cite{Virvillis.2014}\\
\hline
Honeyport & Network port & Server & \ding{55} & \ding{51} & \cite{Fraunholz.2018c}\\
\hline
Decep. web server & Error codes, Robot.txt & Server& \ding{55} & \ding{51}  & \cite{Fraunholz.2018b} \\
\hline
OS interf. & System call & Server & \ding{51} & \ding{55} & \cite{Rowe.2007} \\
\hline
\end{tabular}
\end{table}

\subsection{Honeypots}

\textit{Nawrocki et al.} \cite{Nawrocki.2016} recently reviewed honeypot systems and data analysis.
Therefore this subsection focuses on neglected,
but in the authors' opinions important,
topics in their work.
First,
industrial honeypots are briefly described.
After that,
\textit{Virtual Machine Introspection} (VMI)-based systems are reviewed.
Intelligent honeypots and automated deployment are important topics as well,
but they were extensively reviewed by \textit{Mohammadzadeh et al.} \cite{Mohammadzadeh.2012} and \textit{Zakaria} \cite{Zakaria.2012,Zakaria.2013} in 2012 and 2013 and are therefore not considered in this work.
More recent work in this domain includes adaptive deployment strategies \cite{Fraunholz.2017g,Fraunholz.2017h} and machine learning-based analysis \cite{Fraunholz.2017d}.

\subsubsection{Industrial Honeypots}

An interesting field of application for \textit{HP}s is an industrial environment.
Industrial systems and critical infrastructure are essential for modern societies \cite{DuqueAnton.2017}.
In 2004,
the \textit{Critical Infrastructure Assurance Group} from \textit{Cisco Systems Inc.} published a \textit{HP} especially designed for industrial networks \cite{Pothamsetty.2004}.
Their system supported \textit{Modbus},
\textit{FTP},
\textit{telnet},
\textit{HTTP} and was based on \textit{honeyd} \cite{Provos.2004}.
Four years later,
the first industrial \textit{high-interaction honeypot} (HIHP) was introduced by \textit{Digital Bond.} \cite{DigitalBondInc..2007}.
The most prominent industrial \textit{HP} is \textit{Conpot} \cite{Rist.2015} published by \textit{Rist} in 2013.
\textit{Conpot} is a classified as low-interaction \textit{HP} but supports a vast number of industrial communication protocols such as \textit{Modbus},
\textit{IPMI},
\textit{SNMP},
\textit{S7} and \textit{Bacnet}.
Many industrial honeypots are based on either \textit{honeyd} or \textit{Conpot}.
\textit{Conpot}-based systems are extended to \textit{HIHP}s by combining them with a \textit{Matlab/Simulink}-based simulation of an air conditioning system \cite{Litchfield.2017},
the power grid simulation \textit{gridlabd} \cite{GridPot:SymbolicCyberPhysicalHoneynetFramework.2015, Redwood.2015} and 
the \textit{IMUNES} network simulator \cite{Kuman.2017}.
\textit{Wilhoit and Hilt} \cite{Wilhoit.2015} conducted an experiment with a \textit{HP} imitating a gas station monitoring software.
They captured several attacks against the system e.g. a \textit{DDoS}-attack that they attributed to the \textit{Syrian Electronic Army} (SEA).
It was pointed out by \textit{Winn et al.} \cite{Winn.2015} that cost-effective deployment of \textit{HP} is crucial for their establishment as security mechanisms.
They experimented with \textit{honeyd} for cost-effective deployment of different real world industrial control systems.
Another technique for this was proposed in 2017 by \textit{Guarnizo et al.} \cite{Guarnizo.2017}.
They forwarded incoming traffic from globally deployed so called \textit{wormhole} servers to a limited number of \textit{IoT}-devices.
This technique creates a vast number of deployed systems with only a small number of real hardware devices.

\begin{table}[ht]
\renewcommand{\arraystretch}{1.3}
\caption{Comparison of the different research for industrial \textit{HP}s}
\label{table_industrial}
\centering
\scriptsize
\begin{tabular}{c c|c|c|c|}
\cline{3-5}
& & \multicolumn{3}{c|}{\textbf{Basis HP}}\\
\hhline{|~|~|-|-|-|}
& & \multicolumn{1}{c}{{\cellcolor{lightgray!50}}Conpot} & \multicolumn{1}{|c|}{{\cellcolor{lightgray!50}}Honeyd} & \multicolumn{1}{c|}{{\cellcolor{lightgray!50}}Other}\\
\hline
\multicolumn{1}{|c}{\multirow{2}{*}{\textbf{Interaction}}} & \multicolumn{1}{|c|}{{\cellcolor{lightgray!50}}Low} & \cite{Rist.2015, Kuman.2017, Cao.2018} & \cite{Pothamsetty.2004, Gantsou.2014, Winn.2015} & \cite{Simoes.2013, Holczer.2015, Haney.2014, Vasilomanolakis.2015, Wilhoit.2015, Koltys.2015} \\
\hhline{|~|-|-|-|-|}
\multicolumn{1}{|c}{}& \multicolumn{1}{|c|}{{\cellcolor{lightgray!50}}High} & \cite{Scott.2014, GridPot:SymbolicCyberPhysicalHoneynetFramework.2015, Redwood.2015} & - & \cite{DigitalBondInc..2007, Disso.2013, Buza.2014,PaPa.2015, Antonioli.2016, Guarnizo.2017, Irvene.2018, Litchfield.2017} \\
\hline
\end{tabular}
\end{table}

\textit{Jicha et al.} \cite{Jicha.2016} conducted an in-depth analysis of \textit{Conpot}.
A six months long-term study with conpot was conducted by \textit{Cao et al.} \cite{Cao.2018}.
\textit{Serbanescu et al.} \cite{Serbanescu.2015} in contrast employed a large-scale honeynet offering a variety of different industrial protocols for a 28 day experiment.

\subsubsection{VMI-based High-interaction Honeypot}
The authors believe that VMI-based \textit{HP}s are the most recent and also most future-proof technology for \textit{HIHP}s.
In 2017,
\textit{Sentanoe et al.} \cite{Sentanoe.2017} compared different (HIHP) technologies.
They also decided for \textit{VMI} as a technology for their experiments with \textit{SSH} \textit{HP}s.
\textit{Vrable et al.} \cite{Vrable.2005} introduced the \textit{Potemkin} honeyfarm.
The honeyfarm is based on \textit{Xen} and is able to clone a reference virtual machine for each attack.
\textit{Argos} was published by \textit{Portokalidis et al.} \cite{Portokalidis.2006} in 2006.
It is able to conduct a taint analysis to fingerprint malware.
Later on,
the Honeynet Project \cite{TheHoneynetProject.2010} modified \textit{Sebek} to include \textit{VMscope} \cite{Jiang.2007}.
\textit{VMwatcher} \cite{Jiang.2007b} is able to clone a \textit{VM}.
The cloned \textit{VM} is monitored by \textit{AV}, \textit{IDS} or high-level forensic methods.
Hay and Nance \cite{Hay.2008} proposed another idea for high-level forensic methods by running the forensic tools directly on the \textit{VM} while it is paused.
\textit{Dolan-Gavitt et al.} \cite{DolanGavitt.2011} automated the generation of introspection scripts by translating in-guest applications into out-guest introspection code.
The work of \textit{Pfoh et al.} \cite{Pfoh.2011} focused on performance and flexibility for \textit{QEMU/KVM}-based system call tracing.
\textit{Timescope} \cite{Srinivasan.2011} was proposed in 2011.
It is able to record an intrusion and replay it several times to observe different aspects.
\textit{Biedermann et al.} \cite{Biedermann.2012} published a framework which is able to live-clone a \textit{VM} in case of an occurring attack.
The attacker is then redirected to the cloned instance and system calls as well as network activity and the memory state are monitored.
A hybrid architecture was proposed by \textit{Lengyel et al.} \cite{Lengyel.2012},
they focus on the detection of malware with a combination of \textit{low-interaction honeypots} (LIHPs) and \textit{HIHP}s.
Later on,
they published an advanced version of their architecture with the cloning mechanism from the \textit{Potemkin} honeyfarm \cite{Vrable.2005} and the monitoring mechanism from their previous work.
In 2014,
\textit{Lengyel et al.} \cite{Lengyel.2014},
again,
proposed an advanced version of their framework with additional features for automated deplyoment and malware analysis.
The advantages of \textit{VMI}-based \textit{HP}s for deception systems based on \textit{MTD} was pointed out by \textit{Urias et al.} \cite{Urias.2015}.
\textit{Shi et al.} \cite{Shi.2015} introduced a framework with an integrated module for the analysis of system call traces.

\begin{table}[ht]
\renewcommand{\arraystretch}{1.3}
\caption{Comparison of the different research for VMI-based honeypots}
\label{table_vmi}
\centering
\scriptsize
\begin{tabular}{|L{2.5cm}|L{0.8cm}|L{2.5cm}|L{2cm}|L{3cm}|}
\hline
\rowcolor{lightgray!50}\textbf{Author} & \textbf{Year} & \textbf{Research obj.} & \textbf{Virt. engine}& \textbf{Monitoring}\\
\hline
Vrable et al. \cite{Vrable.2005} & 2005 & Generation & QEMU/KVM & Unknown \\
\hline
Portokalidis et al. \cite{Portokalidis.2006} & 2006 & Signatures & QEMU/KVM & Volatile memory, taint analysis \\
\hline
Jiang and Wang \cite{Jiang.2007} & 2007 & Stealthiness & QEMU/KVM & System calls \\
\hline
Jiang et al. \cite{Jiang.2007b} & 2007 & Semantic gap & VMware, Xen, QEMU/KVM, UML & Full system \\
\hline
Hay and Nance \cite{Hay.2008} & 2008 & High-level forensics & Xen & Emulated Unix utilities \\
\hline
Tymoshyk et al. \cite{Tymoshyk.2009} & 2009 & Semantic gap & QEMU/KVM & System calls \\
\hline
Honeynet Project \cite{TheHoneynetProject.2010} & 2010 & VMI-Sebek & QEMU/KVM & System calls  \\
\hline
Dolan-Gavitt et al. \cite{DolanGavitt.2011} & 2011 & Automation & QEMU/KVM & System calls  \\
\hline
Pfoh et al. \cite{Pfoh.2011} & 2011 & Performance & QEMU/KVM & System calls \\
\hline
Srinivasan and Jiang \cite{Srinivasan.2011} & 2011 & Record and replay & QEMU/KVM & System calls \\
\hline
Biedermann et al. \cite{Biedermann.2012} & 2012 & Generation & Xen & System calls, network monitoring, memory state \\
\hline
Lengyel et al. \cite{Lengyel.2012} & 2012 & Generation, file capturing & Xen & Memory state \\
\hline
Lengyel et al. \cite{Lengyel.2013} & 2013 & Routing & Xen & Memory state \\
\hline
Beham et al. \cite{Beham.2013} & 2013 & Visualization, performance & KVM, Xen & VMI-honeymon \cite{Lengyel.2012} \\
\hline
Lengyel et al. \cite{Lengyel.2014} & 2014 & Automation & Xen & System calls, file system \\
\hline
Urias et al. \cite{Urias.2015} & 2015 & Generation & KVM & System calls, processes, file system \\
\hline
Shi et al. \cite{Shi.2015} & 2015 & System call analysis & KVM & System calls \\
\hline
Sentanoe et al. \cite{Sentanoe.2017} & 2017 & SSH & Unknown & System calls \\
\hline
\end{tabular}
\end{table}

Recently,
the idea of using Linux containers as an alternative for virtual machines was investigated by \textit{Kedrowitsch} \cite{Kedrowitsch.2017}.
They concluded that containers are well suited for the deployment on low-powered devices but are trivial to detect.

\subsubsection{Anti-Honeypot}
As any software,
honeypots are prone to software bugs.
In 2004,
\textit{Krawetz} \cite{Krawetz.2004} introduced the idea to identify honeypots by probing the functionality of the simulated services.
In his work,
he proposed to send e-mails from \textit{SMTP}-\textit{HP}s to himself.
If no mails are received the suggested functionality is not available and the systems' environment is suspicious.
\textit{Fu et al.} \cite{Fu.2006} propose several methods to detect virtual \textit{HP}s such as \textit{honeyd} based on the temporal behavior.
They use Ping, \textit{TCP} and \textit{UDP} based approaches to determine the round trip time of a packet.
\textit{Mukkamala et al.} \cite{Mukkamala.2007} integrated machine learning into the temporal behavior based detection of honeypots.
Counter measures against this type of fingerprinting are developed by \textit{Shiue and Kao} \textit{Shiue.2008}.
They propose \textit{honeyanole} to mitigate fingerprinting based on temporal behavior by traffic redirection.
\textit{Bahram et al.} \cite{Bahram.2010} investigated \textit{VMI} and found that it is prone to changes in the kernel memory layout.
Changes in the layout increase the semantic gap and render \textit{VMI} infeasible.
The issues of missing customization was discussed by \textit{Sysman et al.} \cite{Sysman.2015}.
They used \textit{Shodan} to identify the addresses of over 1000 conpot deployments based on the fictional default company name.
A taxonomy on anti-visualization and anti-debugging techniques was published by \textit{Chen et al.} \cite{Chen.2008}.
They divide these techniques by:
abstraction,
artifact,
accuracy,
access level,
complexity,
evasion and imitation.
Additionally,
they published a remote fingerprinting method to fingerprint virtualized hosts.
Another taxonomy on \textit{HP} detection techniques was published by \textit{Uitto et al.} \cite{Uitto.2017}.
They define temporal,
operational,
hardware and environment as fundamental classes.
In 2016,
\textit{Dahbul et al.} \cite{Dahbul.2016} introduced a threat model for \textit{HP}s.
This model groups attacks in three groups:
poisoning,
compromising and learning.
They also propose a number of fixes for \textit{honeyd},
\textit{Dianaea} and \textit{Kippo}.

\begin{table}[ht]
\renewcommand{\arraystretch}{1.3}
\caption{Overview of counter measures against \textit{HP}s}
\scriptsize
\label{table_anti}
\begin{tabular}{|p{1.5cm}|p{4cm}|p{2cm}|p{3cm}|p{1.2cm}|}
\hline
\rowcolor{lightgray!50}\textbf{Method} & \textbf{Detail} & \textbf{Target} & \textbf{Mitigation} & \textbf{Ref.}\\
\hline
Temporal behavior &	Measure RTT to expose correlations between IP addresses &	honeyd, virtual honeypots &	Simulating timing behavior & \cite{Fu.2006,Chen.2008,Mukkamala.2007,Dornseif.2004,Holz.2005}\\
\hline
Stack fingerprinting &	Send corrupted packets and analyze responses &	Simulated communication stacks &	Implementation of full TCP/IP-stack & \cite{Corey.2003,Morris.2014}\\
\hline
Functional probing &	Use provided functions and verify status & SMTP and DNS &	Implementation of full functionality & \cite{Krawetz.2004, Dornseif.2004} \\
\hline
System call behaviour	& Anomalies in temporal behavior or memory locations &	Linux systems &	Simulating timing behaviour, KASLR & \cite{Dornseif.2004,Holz.2005,Corey.2003,Corey.2003b} \\
\hline
Network traffic &	Analyze RX and TX network traffic e.g. number of bytes & Network based data exfiltration e.g. Sebek	 &Hinder network monitoring, VMI, Proxy & \cite{Dornseif.2004}\\
\hline
UML detection &	dmesg output, network device, /proc/, memory laylout & UML based host isolation &	Manipulating tools to show related information & \cite{Holz.2005}\\
\hline
VMware detection & Hardware e.g. MAC address, I/O backdoor & VMware based host isolation & Customize hardware, patch I/O backdoor \cite{Corey.2003} & \cite{Holz.2005,Corey.2003}\\
\hline
Debugger detection & Use ptrace() function, IsDebuggerPresent() function or memory search for 0xCC & e.g. Cuckoo & - & \cite{Ferrand.2013,Holz.2005} \\
\hline
Semantic gap & Manipulate kernel data structure	& VMI & - & \cite{Bahram.2010} \\
\hline
Customiz. & Search for default strings & - & Customize systems & \cite{Sysman.2015, Dahbul.2016, Dornseif.2004, Holz.2005,Corey.2003} \\
\hline
\end{tabular}
\end{table}

\textit{Hayatle et al.} \cite{Hayatle.2012} proposed a method based on \textit{Dempster-Shafer} evidence combining to use multi-factor decision making to detect \textit{HP}s.
Later on,
they published a work on the detection of \textit{HIHP}s based on \textit{Markov} models \cite{Hayatle.2013}.
Honeypot-aware attackers or botnets are investigated by \textit{Wang} \cite{Wang.2010} and \textit{Costarella et al.} \cite{Costarella.2015}.
The risks of reflected attacks by the abuse of \textit{HP}s was discussed by \textit{Husak} \cite{Husak.2013}.

\subsection{Field studies and technological surveys}
\label{ssec:benchmarks}
In this subsection,
field studies and technological surveys are considered.
First, 
applications and deployments are examined.
The kind and duration of deployment was analyzed,
as well as the \textit{DT},
success,
attack vector and format of retrieved data were part of the investigation.
After that,
works that evaluate deception resources are analyzed.
They are grouped according to the deception resources they consider,
as well as metrics and characteristics that are considered by the authors.
A summary of technological surveys of deception resources is given in table~\ref{tab:benchmarks}.
14 noteworthy examples of field studies about deception technologies were found.
\textit{Cohen et al.}~\cite{Cohen.2001} performed an analysis of real attackers' behavior in a vulnerable system.
They introduced security experts to a system under attack in which they had deployed deception resources and monitored their behavior.
Results were derived from a questionnaire answered by the experts after the experiment.
\textit{Fraunholz et al.} performed two field studies.
In the first study,
\textit{Fraunholz and Schotten}~\cite{Fraunholz.2018b} proposed server-based deception mechanisms in order to hinder attackers.
Fake banners, 
fake \textit{Robots.txt},
tampered error response,
an adaptive delay and honey files were presented in order to study attacker behavior under these circumstances.
1200 accesses were monitored.
In the second study~\cite{Fraunholz.2017f,Fraunholz.2017i},
they analyzed attacker behavior monitored by six honeypots deployed in one consumer and five web hosting servers,
during a period of 222 days.
Almost 12 million access attempts were monitored by the \textit{LIHP}s used.
Common protocols, 
such as \textit{HTTP},
\textit{HTTPS},
\textit{FTP},
\textit{POP3},
\textit{SMTP},
\textit{SSH} and \textit{Telnet} were offered by the honeypots.
In addition to that,
industrial protocols \textit{Bacnet},
\textit{Modbus} and \textit{S7} were emulated in order to derive insight about the threat landscape for industrial applications.
\textit{Lazarov et al.}~\cite{Lazarov.2016} purposely leaked forged confidential information in \textit{Google} spreadsheets.
\textit{IP} addresses were contained in these spreadsheets and were supposed to lure attackers.
174 clicks were monitored,
as well as 44 visits by 39 unique \textit{IP}-addresses.
\textit{Liu et al.}~\cite{Liu.2016} followed a similar approach of publishing apparently confidential information.
\textit{SSH} keys were leaked on \textit{github},
luring attackers to connect to \textit{Cowrie}-based \textit{HP}s.
About 31000 unique passwords were monitored,
as well as the user behavior after log in over the duration of two weeks.
\textit{Zohar et al.}~\cite{Zohar.2016} set up a comprehensive organizational network consisting of users,
mail data, 
documents, 
browser profiles and other \textit{IT} resources.
Traps and decoys were introduced into this network.
Similar to the work of \textit{Cohen et al.}~\cite{Cohen.2001},
52 security professionals took a \textit{Capture the Flag} (CTF) challenge in this network and tried to compromise it.
The goal of this experiment was to determine the best suited means for different organizational networks,
as well as the best deployment strategies for traps and decoys in computer networks.
\textit{Howell}~\cite{ChristianJordanMichaelHowell.2011} used \textit{Sebek} in order to derive insight about attacker behavior after compromise.
The dataset of Jones~\cite{Jones.2014},
gathered by \textit{HIHP}s,
was used for this experiment,
containing 1548 accesses by 478 attackers.
\textit{Sobesto}~\cite{BertrandSobesto.2015} analyzed attacker's reactions to system configuration and banners.
Unused \textit{IP} addresses of an university network were used to deploy \textit{Dionaea} honeypots with a number of vulnerabilities and monitor attacker behavior after intrusion.
This took place from 17th of May to 31st of October,
whereby 624 sessions were monitored with the honeypot tool \textit{Spy}.
\textit{Maimon et al.}~\cite{Maimon.2014} performed two runs of their experiments:
One for two and one for six months.
During that time,
86 and 502 computers respectively were set up and made available from the internet.
Upon connection,
some presented warning banners and some didn't.
They contained different vulnerable entry points,
created with \textit{Sebek} and \textit{OpenVZ} as a gateway,
that attracted 1058 and 3768 trespassing incidents respectively.
The aim was to determine the effect of warnings on attackers.
\textit{Kheirkhak et al.}~\cite{Kheirkhah.2013} analyzed the credentials used by login attempts.
Eight \textit{HP}s with enabled \textit{SSH} connectivity were introduced to six different university campus networks over a duration of seven weeks.
98180 connections from 1153 unique \textit{IP}s in 79 countries were captured.
\textit{Zhan et al.}~\cite{Zhan.2013} used different kinds of \textit{HP}s,
namely \textit{Dionaea},
\textit{Mwcollector},
\textit{Amun} and \textit{Nepenthes}.
These were used to perform statistical analysis about attack patterns.
Five periods were distinguished,
during each of which 166 attackers attacked the vulnerable services:
\textit{SMB},
\textit{NetBIOS},
\textit{HTTP},
\textit{MySQL} and \textit{SSH}.
\textit{Salles-Loustau et al.}~\cite{SallesLoustau.2011} analyzed the attacker behavior based on keystroke patterns.
Three honeypots with different configurations captured 211 attack sessions during a period of 167 days.
\textit{Berthier et al.}~\cite{Berthier.2009} focused on the behavior and actions of an attacker after compromising the system.
In order to do so,
24 different actions were monitored as indicators for different types of behavior.
The \textit{HP}s were set up for a duration of eight months in university networks,
were available via \textit{SSH} and captured 20335 typed commands in 1171 attack sessions.
\textit{Ramsbrock et al.}~\cite{Ramsbrock.2007} followed a similar approach.
Four Linux \textit{HP}s were introduced into a university network,
available via \textit{SSH} with easily guessable credentials for a duration of 24 days.
Attacker actions were monitored with \textit{syslog-ng} for capturing commands,
\textit{strace} to log system calls and \textit{Sebek} to collect keystrokes.
269262 attacking attempts from 229 unique \textit{IP}s were collected.
It can be seen that a strong focus in the application of deception technologies lies on \textit{HP}s.
Apart from that,
real systems that are extended with monitoring capacities are often employed.
By definition,
they count as \textit{HIHP}s.
Only two of the works described above employ significantly different \textit{DT}s,
namely \textit{Fraunholz and Schotten}~\cite{Fraunholz.2018b} and \textit{Lazarov et al.}~\cite{Lazarov.2016}.
Seven noteworthy examples of technological surveys about \textit{DT}s were found.
They are summarized in table~\ref{tab:benchmarks}.
In these surveys,
eleven evaluation features that were shared by at least two surveys were identified.
They are numbered from F1 to F11 and are defined as follows:
Interactivity (F1),
scalability (F2),
legal or ethical considerations (F3),
type (F4),
deployment (F5),
advantages and disadvantages in comparison with other kinds of defense technologies (F6),
quality and type of data and the derived insights (F7),
type of the \textit{DT} resource (F8),
technical way of deployment and kind of \textit{DT} (F9),
detectability and anti-detection capabilities (F10) and extensibility (F11).

\begin{table*}[!h]
\centering
\scriptsize
\caption{Overview of technological survey papers}
\label{tab:benchmarks}
\begin{tabular}{| l | c | l | c | c | c | c | c | c | c | c | c | c | c | c |}
\hline
\rowcolor{lightgray!50}\textbf{Work} & \textbf{Year} & \textbf{DT} & \textbf{F1} & \textbf{F2} & \textbf{F3} & \textbf{F4} & \textbf{F5} & \textbf{F6} & \textbf{F7} & \textbf{F8} & \textbf{F9} & \textbf{F10} & \textbf{F11} & \textbf{other}  \\
\hline
\textit{Smith}~\cite{Smith.2016} & 2016 & HP/HT & \ding{51} & \ding{55} & \ding{51} & \ding{51} & \ding{55} & \ding{55} & \ding{51} & \ding{51} & \ding{51} & \ding{51} & \ding{55} & - \\
\hline
\textit{Nawrocki et al.}~\cite{Nawrocki.2016} & 2016 & HP & \ding{51} & \ding{55} & \ding{55} & \ding{51} & \ding{55} & \ding{51} & \ding{51} & \ding{51} & \ding{51} & \ding{55} & \ding{55} & - \\
\hline
\textit{Grudziecki et al.}~\cite{Grudziecki.2012} & 2012 & HP & \ding{51} & \ding{51} & \ding{55} & \ding{51} & \ding{51} & \ding{51} & \ding{51} & \ding{55} & \ding{51} & \ding{55} & \ding{51} & Rel./Sup. \\
\hline
\textit{Girdhar and Kaur}~\cite{Girdhar.2012} & 2012 & HP & \ding{51} & \ding{55} & \ding{55} & \ding{51} & \ding{55} & \ding{51} & \ding{55} & \ding{55} & \ding{55} & \ding{55} & \ding{55} & - \\
\hline
\textit{Gorzelak et al.}~\cite{Gorzelak.2011} & 2011 & HP & \ding{55} & \ding{51} & \ding{55} & \ding{55} & \ding{51} & \ding{55} & \ding{51} & \ding{55} & \ding{51} & \ding{55} & \ding{51} & - \\
\hline
\textit{Lakhani}~\cite{Lakhani.2003} & 2003 & HP & \ding{55} & \ding{55} & \ding{51} & \ding{51} & \ding{51} & \ding{55} & \ding{55} & \ding{55} & \ding{51} & \ding{55} & \ding{55} & - \\
\hline
\end{tabular}
\end{table*}

\textit{Smith}~\cite{Smith.2016} considers a variety of \textit{DT}s.
\textit{HP}s, 
as well as \textit{HT}s and honeycreds,
honeytraps and honeynets are examined as general concepts.
In terms of tools,
he analyzes \textit{VMWare},
\textit{Chroot} and \textit{Honeyd} with respect to their usability as \textit{DT} tools.
\textit{Nawrocki et al.}~\cite{Nawrocki.2016} analyze and compare 68 different \textit{HP}s.
\textit{Grudziecki et al.}~\cite{Grudziecki.2012} analyze 33 different \textit{HP}s.
They are the only ones considering reliability and support of the tools as an evaluation feature.
\textit{Bringer et al.}~\cite{Bringer.2012} survey a number of more than 80 paper that present \textit{HP}s technologies,
60 of which supposedly had a significant impact on the field of \textit{DT}.
\textit{Girdhar and Kaur}~\cite{Girdhar.2012} analyze five different \textit{HP}s:
\textit{ManTrap},
\textit{Back officer friendly},
\textit{Specter},
\textit{Honeyd} and \textit{Honeynet}.
\textit{Gorzelak et al.}~\cite{Gorzelak.2011} compare honeypots with other incident detection and response tools.
30 different tools or services and twelve different methodologies are considered in their work.
\textit{Lakhani}~\cite{Lakhani.2003} compares four different honeypots in his work:
\textit{LaBrea},
\textit{Specter},
\textit{Honeyd} and \textit{ManTrap}.
In summary,
most works distinguished between type of \textit{HP},
namely research and production.
The two second most important evaluation features are the kind of deception technology and the quality and type of data generated by the resource.
Only \textit{Grudziecki et al.}~\cite{Grudziecki.2012} considered reliability and the technical support of a \textit{DT}.

\section{Legal Considerations and Ethics} \label{sec:legal}

This sections discusses legal and ethical aspects.
\textit{Entrapment}, \textit{privacy} and \textit{liablity} are found to be discussed in most literature.
Therefore table \ref{tab:ethical} discriminates the reviewed literature by these subjects.
Furthermore,
the works are grouped by the country they consider and if ethical aspects are also taken into account.

\textit{Spitzner} \cite{Spitzner.2003} and \textit{Mokube et al.} \cite{Mokube.2007} discussed the \textit{Fourth Amendment to the US Constitution},
while the \textit{Wiretap Act} was analyzed \textit{Burstein} \cite{Burstein.2008},
\textit{Ohm et al.} \cite{Ohm.2007} or \cite{Spitzner.2003}.
\textit{Dornseif et al.} \cite{Dornseif.2004b} discussed aspects of the \textit{Criminal Law} and the \textit{Tort Law}.
The \textit{Patriot Act} was examined \textit{Spitzner} \cite{Spitzner.2003}.
\textit{Burstein} \cite{Burstein.2008} links the \textit{Digital Millennium Copyright Act},
the \textit{Computer Fraud and Abuse Act} and the \textit{Stored
Communications Act}.
\textit{Scottberg et al.} \cite{Scottberg.2002} and \textit{Jain} \cite{Jain.2011} concluded that only law enforcement officers can entrap anyone.
Furthermore,
\textit{Jain} concludes that ``organizations or educational institutions cannot be charged with entrapment''. 
Privacy is granted by the European convention on Human Rights \cite{Ohm.2007} and addressed in the context of deception by \textit{Jain} \cite{Jain.2011},
\textit{Mokube} \cite{Mokube.2007},
\textit{Schaufenbuel} \cite{Schaufenbuel.2008} and \textit{Sokol} \cite{Sokol.2014,Sokol.2017}.
\textit{Fraunholz et al.} \cite{Fraunholz.2017b} adresses privacy,
copyright,
self-defence and domain specific law,
such as rights of law enforcement,
research institutes and telecommunication providers. 
\textit{Liability} is covered in \textit{Schaufenbuel} \cite{Schaufenbuel.2008},
\textit{Campell} \cite{Campbell.2014},
\textit{Fraunholz et al.} \cite{Fraunholz.2017b} and \textit{Nawrocki et al.} \cite{Nawrocki.2016}. 
\textit{Schaufenbuel} \cite{Schaufenbuel.2008} proposed suggestions for the operation and use of \textit{DT}s to mitigate legal risks. 
The 16 works covering aspects of \textit{US} jurisdiction in contrast to the five ones covering European law show a comparatively higher scientific interest in \textit{US} law.
Finally,
\textit{Campbell} \cite{Campbell.2014} is comparing the \textit{US} with the African 
jurisdiction,
while \textit{Warren et. al.} \cite{Warren.2003} addresses Australian law.

\begin{table*}[!h]
\centering
\scriptsize
\caption{An overview of legal and ethical studies on \textit{DT}}
\label{tab:ethical}
\begin{tabular}{ c | c | c | c | c | c | c | c | c }
\hhline{|~|-|-|-|-|-|-|~|~|}
& \multicolumn{3}{c|}{{\cellcolor{lightgray!50}}\textbf{Country}}&\multicolumn{3}{c|}{{\cellcolor{lightgray!50}}\textbf{Legal aspects}} & \multicolumn{1}{c}{}\\
\hline
\rowcolor{lightgray!50}\multicolumn{1}{|l|}{\textbf{Work}} & \textbf{US} & \textbf{Europe}& \textbf{Others} & \textbf{Entrapment} & \textbf{Privacy} & \textbf{Liability} & \multicolumn{1}{l|}{\textbf{Ethics}}\\
\hline
\multicolumn{1}{|l|}{Dornseif \cite{Dornseif.2004b}}&\ding{51} &\ding{55} &\ding{55} &\ding{55} &\ding{51} &\ding{55}&\multicolumn{1}{c|}{\ding{51}}\\
\hline
\multicolumn{1}{|l|}{Spitzner  \cite{Spitzner.2003}}&\ding{51} &\ding{55} &\ding{55} &\ding{51} &\ding{55} &\ding{55}&\multicolumn{1}{c|}{\ding{55}}\\
\hline
\multicolumn{1}{|l|}{Scottberg \cite{Scottberg.2002}}&\ding{51} &\ding{55} &\ding{55} &\ding{51} &\ding{51} &\ding{55}&\multicolumn{1}{c|}{\ding{55}}\\
\hline
\multicolumn{1}{|l|}{Jain \cite{Jain.2011}}&\ding{51} &\ding{55} &\ding{55} &\ding{51} &\ding{51} &\ding{55}&\multicolumn{1}{c|}{\ding{55}}\\
\hline
\multicolumn{1}{|l|}{Mokube \cite{Mokube.2007}}&\ding{51} &\ding{55} &\ding{55} &\ding{51} &\ding{51} &\ding{51}&\multicolumn{1}{c|}{\ding{55}}\\
\hline
\multicolumn{1}{|l|}{Bringer \cite{Bringer.2012}}&\ding{51} &\ding{55} &\ding{55} &\ding{55} &\ding{55} &\ding{55}&\multicolumn{1}{c|}{\ding{51}}\\
\hline
\multicolumn{1}{|l|}{Schaufenbuel \cite{Schaufenbuel.2008}}&\ding{51} &\ding{55}&\ding{55}&\ding{51} &\ding{51} &\ding{51}&\multicolumn{1}{c|}{\ding{55}}\\
\hline
\multicolumn{1}{|l|}{Sokol \cite{Sokol.2014}}&\ding{55} &\ding{55} &\ding{55} &\ding{55} &\ding{51} &\ding{55}&\multicolumn{1}{c|}{\ding{55}}\\
\hline
\multicolumn{1}{|l|}{Campbell \cite{Campbell.2014}}&\ding{55} &\ding{55} &African vs. US &\ding{51}&\ding{51}&\ding{51}&\multicolumn{1}{c|}{\ding{51}}\\
\hline
\multicolumn{1}{|l|}{Nawrocki \cite{Nawrocki.2016}}&\ding{51} &\ding{51} &\ding{55} &\ding{51} &\ding{51} &\ding{51}&\multicolumn{1}{c|}{\ding{55}}\\
\hline
\multicolumn{1}{|l|}{Rowe \cite{Rowe.2016}}&\ding{51} &\ding{55} &\ding{55} &\ding{51} &\ding{55}&\ding{55}&\multicolumn{1}{c|}{\ding{51}}\\
\hline
\multicolumn{1}{|l|}{Burstein \cite{Burstein.2008}}&\ding{51} &\ding{55} &\ding{55} &\ding{55} &\ding{55} &\ding{55}&\multicolumn{1}{c|}{\ding{55}}\\
\hline
\multicolumn{1}{|l|}{Radcliffe \cite{Radcliffe.2007}}&\ding{51} &\ding{55} &\ding{55} &\ding{55} &\ding{55} &\ding{55}&\multicolumn{1}{c|}{\ding{55}}\\
\hline
\multicolumn{1}{|l|}{Rubin \cite{Rubin.2006}}&\ding{51} &\ding{55} &\ding{55} &\ding{55} &\ding{51} & \ding{51}&\multicolumn{1}{c|}{\ding{55}}\\
\hline
\multicolumn{1}{|l|}{Karyda \cite{Karyda.2007}}&\ding{51} &\ding{55} &\ding{55} &\ding{55} &\ding{55} &\ding{55} &\multicolumn{1}{c|}{\ding{55}}\\
\hline
\multicolumn{1}{|l|}{Belloni \cite{Belloni.2015}}&\ding{51} &\ding{55} &\ding{55} &\ding{55} &\ding{55} &\ding{55} &\multicolumn{1}{c|}{\ding{55}}\\
\hline
\multicolumn{1}{|l|}{Sokol \cite{Sokol.2017}}&\ding{55} &\ding{51} &\ding{55} &\ding{55} &\ding{51} &\ding{55} &\multicolumn{1}{c|}{\ding{55}}\\
\hline
\multicolumn{1}{|l|}{Ohm \cite{Ohm.2007}}&\ding{55} &\ding{51} &\ding{55} &\ding{55} &\ding{51} &\ding{51} &\multicolumn{1}{c|}{\ding{55}}\\
\hline
\multicolumn{1}{|l|}{Nance \cite{Nance.2011}}&\ding{51} &\ding{55} &\ding{55} &\ding{55} &\ding{55} &\ding{55} &\multicolumn{1}{c|}{\ding{51}}\\
\hline
\multicolumn{1}{|l|}{Warren \cite{Warren.2003}}&\ding{55} &\ding{55} &Australia &\ding{55} &\ding{55} &\ding{55} &\multicolumn{1}{c|}{\ding{55}}\\
\hline
\multicolumn{1}{|l|}{Dornseif \cite{Dornseif.2004b}}&\ding{55} &\ding{51} &\ding{55} &\ding{55} &\ding{55} &\ding{55} &\multicolumn{1}{c|}{\ding{51}}\\
\hline
\multicolumn{1}{|l|}{Fraunholz et al. \cite{Fraunholz.2017b}}&\ding{55} &\ding{51} &\ding{55} &\ding{51} &\ding{51} &\ding{51} &\multicolumn{1}{c|}{\ding{55}}\\
\hline
\end{tabular}
\end{table*}

In contrast to the 22 works covering the legal aspects,
there are only six works addressing ethical issues with \textit{DT}s as well.
\textit{Dornseif et al.} \cite{Dornseif.2004b} describe the problem of making the internet safer by introducing vulnerabilities to the public internet.
The question how it is possible to secure the internet by adding weaknesses,
and about the moral soundness of making people part of an experiment without their knowledge and consent is asked by \textit{Holz} \cite{Holz.2004}.
\textit{Campbell} \cite{Campbell.2014} questions moral implications of enticing someone to commit a crime,
as well as mentions the problem of ``adding fuel to the fire'' when making a system vulnerable to attacks.
\textit{Rowe and Rrushi} \cite{Rowe.2016} attribute the ethical responsibility of \textit{DT} to the programmer,
while addressing the issue of self-modifying code and artificial intelligence.

Inclusion of \textit{DT} into guidelines is only just starting.
The \textit{German federal office for information security} (BSI) published a baseline protection guideline \cite{GermanFederalOfficeofInformationSecurity.2018},
which mentions the use of \textit{HP}s.
They consider them as anomaly detection mechanisms,
while not mentioning \textit{DT}s explicitly.
Implicitly,
however,
they are described,
for example in spoofing server banners.

\section{Conclusion} \label{sec:conclusion}

In this work,
the current state of the art in \textit{DT} and adjacent domains is presented.
Where applicable,
previous surveys are referenced and supplemented with recent research.
It was pointed out that \textit{DT} is a beneficial extension for traditional \textit{IT}-security.
Emphasis was placed on requirement categories,
such as psychological,
formal,
legal and ethical,
as well as on recent trends,
such as \textit{VMI} and the field of industrial and critical infrastructure security.

\section*{Acknowledgment}
This work has been supported by the Federal Ministry of Education and Research of the Federal Republic of Germany (Foerderkennzeichen KIS4ITS0001, IUNO). The authors alone are responsible for the content of the paper.

\bibliographystyle{unsrt}
\bibliography{bibi}

\end{document}